\documentclass[aps,twocolumn,prb,showpacs,amssymb]{revtex4}
\usepackage{graphicx}
\usepackage{amsmath}

\newcommand{\bra}[1]{\langle #1 \lvert}
\newcommand{\ket}[1]{\lvert #1 \rangle}
\newcommand{\G}{\Gamma}

\newcommand{\eps}{\epsilon}
\renewcommand{\t}{\tau}
\renewcommand{\a}{\alpha}
\newcommand{\s}{\sigma}
\newcommand{\p}{\pi}
\renewcommand{\t}{\tau}
\renewcommand{\l}{\ell}

\newcommand{\phd}{\phantom{\dagger}}

\begin{document}

\title{A benzene interference single-electron transistor}

\author{D.~Darau, G.~Begemann, A.~Donarini, and M.~Grifoni}

\affiliation{Institut f\"{u}r Theoretische Physik, Universit\"at
Regensburg, 93035 Regensburg, Germany}

\date{\today}

\begin{abstract}
Interference effects strongly affect the transport characteristics of a
benzene single-electron transistor (SET) and for this reason we call it
interference SET (I-SET). We focus on the effects of degeneracies between
many-body states of the isolated benzene. We show that the particular current
blocking and selective conductance suppression occurring in the benzene I-SET
are due to \emph{interference} effects between the orbitally
degenerate states. Further we study the impact of reduced symmetry
due to anchor groups or potential drop over the molecule. We
identify in the \emph{quasi-degeneracy} of the involved molecular
states the necessary condition for the robustness of the results.
\end{abstract}

 \maketitle

%///////////////////////////////////////////////////////////////////////////////
\section{Introduction}
%///////////////////////////////////////////////////////////////////////////////

Molecular electronics, due to perfect reproducibility and
versatile chemical tailoring of its basic components, represents
one of the most promising answers to the increasing
miniaturization demand of information technology. A crucial issue
in molecular electronics is thus the understanding of the
conduction characteristics through single molecules
\cite{Cuniberti-book}.

Single-molecule-transport measurements rely on the fabrication of
a nanogap between source and drain electrodes and the formation of
a stable molecule-electrode contact. Nanogaps are nowadays
routinely obtained using different techniques including
electromigration
\cite{ParkJ02,Liang02,Kubatkin03,Yu05,Danilov08,Chae06,Poot06,Heersche06,
Osorio07},
mechanical
break-junction\cite{Loertscher07,Smit02,Kiguchi08,Champagne05} and
scanning tunnelling microscopy\cite{Venkataraman06,Repp05,Xiao04}.
Also the challenging goal of effectively gating a nanometer sized
molecule in presence of macroscopic metallic leads has been
achieved \cite{Gittins00,Champagne05,Danilov08}.

A stable contact between molecule and leads is commonly realized
with the mediation of anchor groups attached to the molecule
during its chemical synthesis. Also direct coupling of the
molecule to the electric leads, though, has been very recently
reported\cite{Kiguchi08}. One of the advantages of the first
connecting method is some control over the contact configuration
of the molecule\cite{Mayor03} and the possibility to design the
strength of the tunnelling coupling by choosing specific anchor
groups\cite{Xiao04,Chen06,Danilov08,Venkataraman06b}. All previous
achievements combined with experience accumulated with
semiconducting and carbon-based single electron transistors (SET)
allowed in recent years to measure stability diagrams of single
molecule transistor devices thus realizing molecular spectroscopy
via transport
experiments\cite{ParkJ02,Liang02,Kubatkin03,Yu05,Danilov08,Chae06,Poot06,
Heersche06,Osorio07}.
%%%
\begin{figure}[h!]
  \includegraphics[width=0.9\columnwidth,angle=0]{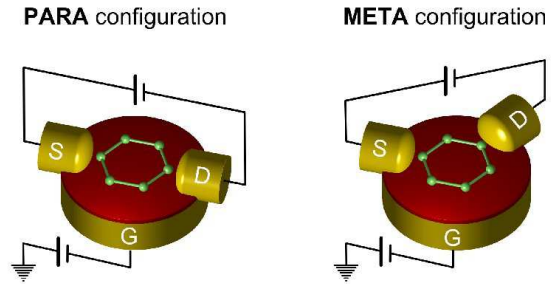}
  \caption{(color online) Schematic representation of the two different
  setups for the benzene I-SET considered in this paper. The
  molecule, lying on a dielectric substrate, is weakly contacted
  to source and drain leads as well as capacitively gated.
  }
  \label{fig1}
\end{figure}
%%%

Single molecule transistors display transport properties which are
very different from those of conventional single electron
transistors.  In fact, vibrational or torsional modes
\cite{Chae06,Osorio07} and intrinsic symmetries of the molecule
can hinder or favor transport through the molecular SET, visible
\textit{e.g.} in the absence or presence of specific excitation
lines in the stability diagram, or in negative differential
conductance features. Many-body phenomena as \textit{e.g.} the
Kondo effect, have been observed as well
\cite{ParkJ02,Liang02,Yu05,Osorio07,Roch08}.

Despite the experimental progress, the theoretical understanding
of the properties of single organic molecules coupled to
electrodes is far from being satisfactory. On the one hand,
numerical approaches to transport based on the combination of
Green's function methods with tight binding model or density functional theory
have become standard in the study of transport at the nanoscale
\cite{Cuniberti-book}. These methods are appropriate to investigate quantum
transport through molecular bridges strongly coupled to leads. In this regime
various groups have recently discussed the possibility to observe interference
effects\cite{Cardamone06,Gagliardi07,Ke08,Qian08}, {\it e.g.} in conjugated
monocyclic
molecules as benzene or annulene \cite{Cardamone06, Ke08}. However, for the
description of transport through a molecule weakly coupled
to leads, other methods are required. In the Coulomb Blockade regime, for
example, due to the crucial role played by the Coulomb interaction in these
systems it is common to resort to a Pauli rate equation \cite{Bruus-book} or to
a generalized master equation for the reduced density matrix (RDM).  For
example, in Hettler \textit{et al.} \cite{Hettler03}, an electronic structure
calculation was performed in order to construct an effective interacting
Hamiltonian for the $\pi$ orbitals of benzene, and the I-V characteristics
of the corresponding molecular junction were calculated within the rate equation
approach.\\
In the presence of degenerate states, however, coherences of the density matrix
influence the dynamics and a master equation approach is appropriate
\cite{Gurvitz96, Braun04, Wunsch05, Donarini06, Mayrhofer07, Koller07,
Begemann08, Schultz09}.
Such coherences can give rise to precession effects in spin transport
\cite{Braun04,Koller07} or cause interference in a molecular single electron
transistor \cite{Donarini06,Koller07,Begemann08}. In the present work we wish to
generalize the discussion on interference phenomena in a benzene
interference-SET presented in \cite{Begemann08} to the case in which the perfect
degeneracy is broken due {\it e.g.}  to contact effects or to the applied
external bias. To this extent the master equation used in \cite{Begemann08} will
be generalized to treat the case of quasi-degenerate
states. Conditions for the persistence of interference phenomena are identified.
We observe that the effects of quasi-degenerate states on transport have been
very recently addressed also in \cite{Schultz09}.
We treat the transport through the benzene I-SET in two different setups, the
para and the meta configuration, depending on the position of the leads with
respect to the benzene molecule (see Fig.~\ref{fig1}). Similar to
\cite{Hettler03}, we start from an interacting Hamiltonian of isolated benzene
where only the localized $p_z$ orbitals are considered and the ions are assumed
to have the same spatial symmetry as the relevant electrons. We calculate the
$4^6=4096$ energy eigenstates of the benzene Hamiltonian numerically.

Subsequently, with the help of group theory, we classify the
eigenstates according to their different symmetries and thus give
a group-theoretical explanation to the large degeneracies
occurring between the electronic states. For example, while the
six-particles ground state ($A_{1g}$ symmetry) is non-degenerate,
there exist four seven-particle ground states due to spin and
orbital ($E_{2u}$ symmetry) degeneracy. Fingerprints of these
orbital symmetries are clearly visible in the strong differences
in the stability diagrams obtained by coupling the benzene I-SET
to the leads in the meta and para configurations. Striking are the
selective reduction of conductance and the appearance of regions of
interference driven current blocking with associated negative
differential conductance (NDC) when changing from the para to the
meta configuration.

NDC and current blocking for benzene junctions have been predicted
also in \cite{Hettler03}, but in the para configuration \emph{and}
in presence of an external electromagnetic field. In our work NDC
occurs despite the absence of an external field in the unperturbed
setup and with no asymmetry in the tunnelling rates. In fact, NDC and current
blocking triggered by interference take place any time a SET
presents an $N$-particle non-degenerate state and two degenerate $N+1$-particle
states such that the ratio between the transition \emph{amplitudes}
$\gamma_{i\alpha}$ ($i=1,2,\quad \alpha=L,R$) between those $N$- and
$N+1$-particle states is different for tunneling at the left ($L$) and at the
right ($R$) lead:
\begin{equation}
\frac{\gamma_{1L}}{\gamma_{2L}} \neq \frac{\gamma_{1R}}{\gamma_{2R}}.
\label{eq:interf_condition}
\end{equation}
Notice that no asymmetry in the tunnelling \emph{rates}, which are proportional
to $|\gamma_{i\alpha}|^2$, is implied by \eqref{eq:interf_condition}. This fact
excludes the interpretation of the physics of the interference SET in terms of
standard NDC with asymmetric couplings. Due to condition (1) there exist linear
combinations of the degenerate $N+1$-particle states which are coupled to one of
the leads but \emph{not} to the other. The state which is decoupled from the
right lead represents a blocking state for the current flowing $ L \to  R$ since
electrons can populate this state by tunnelling from the left lead but cannot
tunnel out towards the right lead. Viceversa the state decoupled from the left
lead is a blocking state for the current $R \to L$. Typically these two blocking
states are not orthogonal and thus cannot form together a valid basis set. The
basis set that diagonalizes the stationary density matrix (what we call in the
manuscript the "physical basis") contains at large positive biases the $L \to R$
blocking state and is thus different from the physical basis at large negative
biases which necessarily contains the $R \to L$ blocking state.
More generally the "physical basis" \emph{depends continuously} on the bias.
Thus only a treatment that includes coherences in the density matrix can
capture the full picture at all biases. By neglecting for simplicity the spin
degree of freedom, the 7-particle ground state of benzene is two
times degenerate while the 6-particle one is non-degenerate. If we choose for
the 7-particle states the eigenstates of the $z$-projection of the angular
momentum we obtain the relation:
%%%
\begin{equation}
\frac{\gamma_{1L}}{\gamma_{2L}}  = \frac{\gamma_{1R}}{\gamma_{2R}} e^{4i\phi},
\label{eq:B_interf_condition}
\end{equation}
%%%
where $\phi$ is the angle between the left and the right lead. Thus in the meta
configuration ($\phi = 2\pi/3$) the condition \eqref{eq:interf_condition} is
fullfilled while in the para ($\phi = \pi$) the amplitude ratios are equal. This
condition implies that, in the para configuration one of the 7-particle states
is decoupled from \emph{both} leads at the same time and can thus (in first
approximation) be excluded from the dynamics.
In contrast, in the meta configuration, the linear combination of uniformly
distributed eigenstates of the angular momentum creates states with a peculiar
interference pattern. The position of their nodes allows to characterize them as
different blocking states.

This paper is outlined as follows: in Section~\ref{theory} we
introduce the model Hamiltonian of the system and present a
density matrix approach setting up a generalized master equation
describing the electron dynamics. We give the expression for the
current in the fully symmetric setup (the generalized master
equation and current formula for the setup under perturbation are
given in Appendix~\ref{appendixGME}). Further we provide a
detailed analysis of the symmetry characteristics of the molecular
eigenstates.\\
In Section~\ref{clean} we present numerical and analytical results
of transport calculations for the unperturbed setup. We study the
occurring interference effects and provide an explanation of the
phenomena based on symmetry considerations.\\
In Section~\ref{ReducedSym} we present the results for the
perturbed setup including a detailed discussion of the transport
in this case. We identify in the \emph{quasi-degeneracy} of the
contributing molecular states the necessary condition for the
robustness of the interference effects.\\
Conclusions and remarks are presented in
section~\ref{conclusions}.
%///////////////////////////////////////////////////////////////////////////////

%///////////////////////////////////////////////////////////////////////////////
\section{Model Hamiltonian and the density matrix approach}\label{theory}
%///////////////////////////////////////////////////////////////////////////////

%-------------------------------------------------------------------------------
\subsection{Model Hamiltonian}\label{model}
%-------------------------------------------------------------------------------

For the description of the benzene molecule weakly coupled to
source and drain leads, we adopt the total Hamiltonian
 $H = H^0_{\rm ben} + H_{\rm leads} + H_{\rm T} + H^\prime_{\rm ben}$.
The first term is the interacting Hamiltonian for isolated
benzene\cite{Pariser53,Pople53,Linderberg68}
\begin{equation}
    \begin{split}
    H^0_{\rm ben} = \,\,
    &\xi_0 \sum_{i\s} d^{\dagger}_{i\s}d^{\phd}_{i\s} +
    b  \sum_{i\s}\left(d^{\dagger}_{i\s}d^{\phd}_{i+1\s} +
d^{\dagger}_{i+1\s}d^{\phd}_{i\s}\right)\\
    +& U \sum_i \left(n_{i\uparrow} - \tfrac{1}{2}\right)
                \left(n_{i\downarrow} - \tfrac{1}{2}\right)\\
    +& V \sum_i\left(n_{i\uparrow} + n_{i\downarrow}- 1\right)
               \left(n_{i+1\uparrow} + n_{i+1\downarrow} -
               1\right),
    \end{split}
    \label{eq:PPP}
\end{equation}
where $d^{\dagger}_{i\sigma}$ creates an electron of spin $\sigma$
in the $p_z$ orbital of carbon $i$, $i = 1,\ldots,6$ runs over the
six carbon atoms of benzene and $n_{i\sigma} =
d^{\dagger}_{i\sigma} d^{\phd}_{i\sigma}$. \\
Only the $p_z$ orbitals (one per carbon atom) are explicitly taken
into account, while the core electrons and the nuclei are combined
into frozen ions, with the same spatial symmetry as the relevant
electrons. They contribute only to the constant terms of the
Hamiltonian and enforce particle-hole symmetry. Mechanical
oscillations are neglected and all atoms are considered
at their equilibrium position.\\
This Hamiltonian for isolated benzene is respecting the $D_{6h}$
symmetry of the molecule. Since for every site there are 4
different possible configurations
$(\ket{0},\ket{\uparrow},\ket{\downarrow},
\ket{\uparrow\downarrow})$, the Fock space has the dimension
$4^6=4096$, which requires a numerical treatment. Though the
diagonalization of the Hamiltonian is not a numerical challenge,
it turns out to be of benefit for the physical understanding of
the transport processes to divide $H_{\rm ben}$ into blocks, according to the
number $N$ of $p_z$ electrons (from 0 to 12), the $z$ projection
$S_z$ of the total spin and the orbital symmetries
of benzene (see Table \ref{table:N6states}).\\
The parameters $b$, $U$, and $V$ for isolated benzene are given in
the literature \cite{Barford-book} and  are chosen to fit optical
excitation spectra. The presence, in the molecular I-SET, of
metallic electrodes and the dielectric is expected to cause a
substantial renormalization of $U$ and $V$ \cite{Kubatkin03,
Kaasbjerg08}. Nevertheless, we do not expect the main results of
this work to be affected by this change. We consider the benzene
molecule weakly coupled to the leads. Thus, to first
approximation, we assume the symmetry of the isolated molecule not
to be changed by the screening. Perturbations due to the
lead-molecule contacts reduce the symmetry in the molecular
junction. They are included in $H^\prime_{\rm ben}$ (see
Eq.~\eqref{eq:H_cont} and \eqref{eq:H_bias}) and will be treated
in
Section~\ref{ReducedSym}.\\
The effect of the gate is included as a renormalization of the
on-site energy $\xi = \xi_0 - eV_{\rm g}$ ($V_{\rm g}$ is the gate
voltage) and we conventionally set $V_{\rm g} = 0$ at the charge
neutrality point. Source and drain leads are two reservoirs of
non-interacting electrons: $H_{\rm leads} =
\sum_{\alpha\,k\,\s}(\eps_k - \mu_{\alpha}) c^{\dagger}_{\alpha k
\s}c^{\phd}_{\alpha k \s}$, where $\alpha = L,R$ stands for the
left or right lead and the chemical potentials $\mu_{\alpha}$ of
the leads depend on the applied bias voltage $\mu_{\rm L,R} =
\mu_0 \pm \tfrac{V_{\rm b}}{2}$. In the following we will measure
the energy starting from the equilibrium chemical potential $\mu_0
= 0$. The coupling to source and drain leads is described by the
tunnelling Hamiltonian
\begin{equation}
 H_{\rm T} = t\sum_{\alpha k \s}
 \left(d^{\dagger}_{\alpha \s} c^{\phd}_{\alpha k \s} +
       c^{\dagger}_{\alpha k \s} d^{\phd}_{\alpha \s}\right),
\end{equation}
where we define $d^{\dagger}_{\alpha \sigma}$ as the creator of an
electron in the benzene carbon atom  which is closest to the lead
$\alpha$. In particular  $d^{\dagger}_{\rm R \sigma} :=
d^{\dagger}_{4 \sigma}, d^{\dagger}_{5 \sigma}$ respectively in
the para and meta configuration, while $d^{\dagger}_{\rm L \sigma}
:= d^{\dagger}_{1 \sigma}$ in both setups.
%-------------------------------------------------------------------------------

%-------------------------------------------------------------------------------
\subsection{Dynamics of the reduced density matrix}\label{dynamics}
%-------------------------------------------------------------------------------

Given the high degeneracy of the spectrum, the method of choice to
treat the dynamics in the weak coupling is the Liouville equation
method already used \textit{e.g.} in \cite{Donarini06,
Mayrhofer07}. In this section we shortly outline how to derive the
equation of motion for the reduced density matrix (RDM) to lowest
non-vanishing order in the tunnelling
Hamiltonian. For more details we refer to \cite{Mayrhofer07, Koller07}.\\
Starting point is the Liouville equation for the total density
operator of molecule and leads $\rho$ in the interaction picture,
treating $H_{\rm T}$ as a perturbation: $i\hbar\frac{d\rho^{\rm
I}(t)}{dt}=\left[H_{\rm T}^{\rm I},\rho^{\rm I}(t)\right]$. This
equation integrated over time and iterated to the second order
reads
\begin{eqnarray}\label{eq:rhoint}
\lefteqn{\dot{\rho}^{\rm I}(t)=-\frac{i}{\hbar}[H^{\rm I}_{\rm
T}(t),\rho^{\rm I}(t_0)]}\\\nonumber &-&\frac{1}{\hbar^2}
\int^t_{t_0} dt^\prime [H^{\rm I}_{\rm T}(t),[H^{\rm I}_{\rm
T}(t^\prime),\rho^{\rm I}(t^\prime)]]
 .\end{eqnarray}
Since we are only interested in the transport through the
molecule, we treat from now on the time evolution of the reduced
density matrix (RDM) $\sigma ={\rm Tr}_{\rm leads}\{\rho^{\rm
I}(t)\}$ \cite{Blum-book}, which is formally obtained from
Eq.~\eqref{eq:rhoint} by tracing out the
leads degrees of freedom: $\dot{\sigma} ={\rm Tr}_{\rm leads}\{\dot{\rho}^{\rm
I}\}$.\\
In order to proceed, we  make the following standard approximations:\\
i) The leads are considered as reservoirs of non-interacting
electrons in thermal equilibrium. Hence we can factorize the
density matrix as $\rho^{\rm I}(t)=\sigma(t)\rho_{\rm s}\rho_{\rm
d}=\sigma(t)\rho_{\rm leads}$.\\
ii) Since the moecule is weakly coupled to the leads we treat the effects of
$H_{\rm T}$ to the lowest non-vanishing order.\\
iii) Due to the continuous interaction of the system with the leads and at high
enough temperature, it is legitimate to apply the Markov approximation and
obtain an equation for $\dot{\sigma}$ which is local in time ($\sigma(t)$
instead of
$\sigma(t^\prime)$ inside the integral). In particular the Markov
approximation becomes exact in the stationary limit $(t \to
\infty)$ we will focus on. Since we are interested in
the long term behavior of the system, we set $t_0\rightarrow
-\infty$ in Eq.~\eqref{eq:rhoint} and finally obtain the generalized master
equation (GME)
\begin{eqnarray}\label{eq:GME2}
\dot{\sigma}(t)= \frac{-1}{\hbar^2}\hspace{-0.2cm}
\int^{\infty}_{0} \hspace{-0.3cm}dt^{\prime\prime} {\rm Tr}_{\rm
leads}\hspace{-0.05cm} \Big\{ [H^{\rm I}_{\rm T}(t),[H^{\rm
I}_{\rm T}(t-t^{\prime\prime}), \sigma(t) \rho_{\rm leads}]]
\Big\}.\nonumber\\
\end{eqnarray}
The reduced density operator $\sigma$ is defined on the Fock space
of benzene, yet we can neglect coherences between states with
different particle number since decoupled from the dynamics of the
populations. For simplicity, we continue here the derivation of
the GME only for the symmetric case with exact orbital degeneracy,
\textit{i.e.}, neglecting $H^\prime_{\rm ben}$
(the perturbed case is presented in Appendix~\ref{appendixGME}).\\
iv) Further we also neglect coherences between states with
different energy (secular approximation). They are irrelevant due to their fast
fluctuation  compared to the dynamics of the system
triggered by the tunnelling coupling.

Under these considerations, it is convenient to express the GME in
terms of the reduced density operator $\sigma^{\rm NE} =
\mathcal{P}_{\rm NE} \,\sigma \,\mathcal{P}_{\rm NE}$, where
$\mathcal{P}_{\rm NE} := \sum_{\ell \t} |N\, E\, \ell\, \t \rangle
\! \langle N\, E\, \ell\, \t |$ is the projection operator on the
subspace of $N$ particles and energy $E$. The sum runs over the
orbital and spin quantum numbers $\ell$ and $\tau$, respectively.
The orbital quantum number $\ell$ distinguishes between orbitally
degenerate states. The exact meaning of $\ell$ will be illustrated
in the next section. In Appendix~\ref{appendixGME}  we derive a GME that retains
coherences also between quasi-degenerate states. That approach
treats with special care the small asymmetries introduced in the
molecule by the coupling to the leads. In fact it interpolates
between the degenerate case treated here and the fully
non-degenerate case in which the GME reduces to a master equation
for populations only.
Equation \eqref{eq:GME2} can be further manipulated by projection into the
subspace of $N$-particle and energy $E$. Since we assume the density matrix to
be
factorized and the leads to be in thermal equilibrium, also the traces over the
leads degree of freedom can be easily performed.
Eventually, the GME for the degenerate case reads
\begin{widetext}
\begin{equation}
    \begin{split}
    \dot{\sigma}^{\rm NE}
    =& -\sum_{\alpha\t} \frac{\G_\alpha}{2}
    \Bigg\{
    \mathcal{P}_{\rm NE}d^{\phd}_{\alpha\t}
    \left[f^+_\alpha(H^0_{\rm ben} - E) - \frac{i}{\pi}\,p_{\alpha}(H^0_{\rm
ben} - E)\right]
    d^{\dagger}_{\alpha\t}\,\sigma^{\rm NE} +\\
    &\phantom{-\sum_{\alpha\t}\frac{\G_\alpha}{2}\,\,\,}
  + \mathcal{P}_{\rm NE}d^{\dagger}_{\alpha\t}
    \left[f^-_\alpha(E - H^0_{\rm ben}) - \frac{i}{\pi}\,p_{\alpha}(E - H^0_{\rm
ben})\right]
    d^{\phd}_{\alpha\t}\,\sigma^{\rm NE} + H.c.
    \Bigg\} + \\
& + \sum_{\alpha\t{\rm E'}} \G_\alpha \mathcal{P}_{\rm NE}
    \Bigg\{
    d^{\dagger}_{\alpha\t} f^+_\alpha(E - E')\, \sigma^{\rm N-1 E'}
d^{\phd}_{\alpha\t}
  + d^{\phd}_{\alpha\t} f^-_\alpha(E' - E)\,\sigma^{\rm N+1
E'}d^{\dagger}_{\alpha\t}
    \Bigg\}
    \mathcal{P}_{\rm NE},
    \end{split}
\label{eq:GME}
\end{equation}
\end{widetext}
where $\Gamma_{\rm L,R}= \tfrac{2\pi}{\hbar}|t_{\rm
L,R}|^2\mathcal{D}_{\rm L,R}$ are the bare transfer rates  with
the constant densities of states of the leads $\mathcal{D}_{\rm
L,R}$. Terms describing sequential tunnelling from and to the lead
$\alpha$ are proportional to the Fermi functions
$f^+_\alpha(x):=f(x - \mu_\alpha)$ and $f^-_\alpha(x) := 1 -
f^+_\alpha(x)$, respectively. Still in the sequential tunnelling
limit, but only in the equations for the coherences, one finds also the
energy non-conserving terms, proportional to the function
$p_{\alpha}(x) = -{\rm Re}\psi\left[\tfrac{1}{2} +
\tfrac{i\beta}{2\pi}(x - \mu_\alpha)\right]$, where $\psi$ is the
digamma function. Both the Fermi functions and the digamma function result
from the trace over the leads degrees of freedom
\cite{Mayrhofer07,Blum-book,Braun04}.\\
A closer analysis of the master equation allows also to formulate
an expression for the current operator. We start from the
definition of the time derivative of the charge on benzene:

\begin{equation}
\frac{d}{dt}\langle Q\rangle={\rm
Tr}\left\{\hat{N}\dot{\sigma}\right\} =\langle\,I_{\rm L}+I_{\rm
R}\,\rangle \label{defI}
\end{equation}
where $Q=\sum_{i\tau}(d^{\dagger}_{i\tau}d^{\phd}_{i\tau}-6)$ is
the operator of the charge on benzene, $\hat{N}$ is the particle
number operator and $I_{\rm L,R}$ are the current operators at the
left(right) contact. Conventionally, in the definition of $I_{\rm
L,R}$ we assume the current to be positive when it is increasing
the charge on the molecule. Thus, in the stationary limit,
$\langle\,I_{\rm L}+I_{\rm R}\,\rangle$ is zero. We write this
expression in the basis of the subspaces of $N$ particles and
energy $E$:
\begin{equation}
\langle\,I_{\rm L}+I_{\rm R}\,\rangle=\sum_{\rm NE}{\rm Tr}
\left\{\hat{N}\mathcal{P}_{\rm NE}\dot{\sigma}\mathcal{P}_{\rm
NE}\right\} =\sum_{\rm NE}{\rm Tr}\left\{N\dot{\sigma}^{\rm
NE}\right\} \label{eq:defI_NE}.
\end{equation}
Further we insert (\ref{eq:GME}) in (\ref{eq:defI_NE}) and take
advantage of the cyclic properties of the trace to find :
\begin{widetext}
\begin{equation}
\begin{split}
 \langle\,I_{\rm L}+I_{\rm R}\,\rangle =
 \sum_{\rm NE}\sum_{\alpha\t}N\Gamma_\alpha{\rm Tr}
 \Bigg\{
 &-\left[
 f^+_\alpha(H^0_{\rm ben}-E)d^\dagger_{\alpha\t}
 \sigma^{\rm NE}d^{\phd}_{\alpha\t}+
 f^-_\alpha(E-H^0_{\rm ben})d^{\phd}_{\alpha\t}
 \sigma^{\rm NE}d^\dagger_{\alpha\t}
 \right]+\\
 & + \sum_{\rm E'}\mathcal{P}_{\rm NE}
 \left[
 f^+_\alpha(E-E')d^\dagger_{\alpha\t}
 \sigma^{\rm N-1E'}d^{\phd}_{\alpha\t}+
 f^-_\alpha(E'-E)d^{\phd}_{\alpha\t}
 \sigma^{\rm N+1E'}d^\dagger_{\alpha\t}
 \right]
 \Bigg\}.
\end{split}
\label{eq:I_Eprime}
\end{equation}
\end{widetext}
Notice that the energy non-conserving contributions drop from the
expression of the current. Still they contribute to the average
current via the density matrix. Since $E$ and $E'$ are dummy
variables, we can switch them in the summands containing $E'$.
Applying the relation:
\begin{equation}
\sum_{\rm NE'}{\rm Tr}\left\{\mathcal{P}_{\rm NE'}\,g(E')\right\}
={\rm Tr}\left\{g(H^0_{\rm ben})\right\},\nonumber
\end{equation}
where $g(E')$ is a generic function, we substitute $E'$ with
$H^0_{\rm ben}$ in Eq.~\eqref{eq:I_Eprime}. Further we can
conveniently rearrange the sum over $N$, arriving at the
expression for the current:

\begin{equation}
    \begin{split}
    \langle\,&I_{\rm L}+I_{\rm R}\,\rangle =
    \sum_{\rm NE}\sum_{\alpha\t}\Gamma_\alpha
    {\rm Tr}\\
    \Bigg\{
        &d^\dagger_{\alpha\t}\sigma^{\rm NE}d^{\phd}_{\alpha\t}
        \!\Big[
            \!-\!Nf^+_\alpha(H^0_{\rm ben}\!-\!E)
            \!+\!(N\!+\!1)f^+_\alpha(H^0_{\rm ben}\!-\!E)
        \Big]\\
        +&d^{\phd}_{\alpha\t}\sigma^{\rm NE} d^\dagger_{\alpha\t}
        \Big[
            \!-\!Nf^-_\alpha(E\!-\!H^0_{\rm ben})
            \!+\!(N\!-\!1)f^-_\alpha(E\!-\!H^0_{\rm ben})
        \Big]\!
    \Bigg\}.
    \end{split}
%\label{eq:I}
\end{equation}

This relation can be further simplified in order to identify the
current operators. The one corresponding to the left contact is
\textit{e.g.}
\begin{equation}
    \begin{split}
    I_{\rm L} = \Gamma_{\rm L}\sum_{\rm NE\tau}\mathcal{P}_{\rm NE}
    \Big[
    &d^{\phd}_{{\rm L}\tau}f^+_{\rm L}(H^0_{\rm ben}-E)d^{\dagger}_{{\rm L}\tau}
    +\\
    &-d^{\dagger}_{{\rm L}\tau}f^-_{\rm L}(E-H^0_{\rm ben})d^{\phd}_{{\rm
L}\tau}
    \Big]\mathcal{P}_{\rm NE}.
    \end{split}
\end{equation}
With this relation we can calculate the stationary current as the
average $\langle I_{\rm L}\rangle ={\rm Tr}\{\sigma_{\rm stat}
I_{\rm L}\} = -\langle I_{\rm R}\rangle$, with $\sigma_{\rm stat}$
as the stationary density operator. The expression of the current
operator for the perturbed system is given in
Appendix~\ref{appendixGME}.
%-------------------------------------------------------------------------------

%-------------------------------------------------------------------------------
\subsection{Symmetry of the benzene eigenstates}\label{Symmetry}
%-------------------------------------------------------------------------------
In this section, we will review the symmetry characteristics of the
eigenstates of the interacting Hamiltonian of benzene, focusing on
the symmetry operations $\s_{\rm v}$ and $C_{n}$ which have a
major impact on the electronic transport through the molecular
I-SET. Benzene belongs to the $D_{6h}$ point group. Depending on
their behavior under symmetry operations, one can classify the
molecular orbitals by their belonging to a certain irreducible
representation of the point group.
\begin{table}[h!]
\centering
\begin{tabular}{cccc|c}
$N$ & \mbox{\hspace{0.2cm}}$\#\uparrow$\mbox{\hspace{0.2cm}}
    & \mbox{\hspace{0.2cm}}$\#\downarrow$\mbox{\hspace{0.2cm}}
    &\mbox{\hspace{0.2cm}}\# states\mbox{\hspace{0.2cm}}
    &\mbox{\hspace{0.2cm}}\# states with a\\
&  &  &  & \mbox{\hspace{0.2cm}}certain symmetry\\
\hline
 6 & 6 & 0 & 1  & 1 $B_{1u}$\\\hline
   &   &   &    & 4 $A_{1g}$\\
& & & & 2 $A_{2g}$\\
& 5 & 1 & 36 &2$\times$6 $E_{2g}$\\
&   &   &    & 4 $B_{1u}$\\
& & & & 2 $B_{2u}$\\
&  &  & &2$\times$6 $E_{1u}$\\\hline
&   &   &    & 16 $A_{1g}$\\
& & & & 20 $A_{2g}$\\
& 4 & 2 & 225 &2$\times$36 $E_{2g}$\\
&   &   &    & 22 $B_{1u}$\\
& & & & 17 $B_{2u}$\\
&  &  & &2$\times$39 $E_{1u}$\\\hline
&   &   &    & 38 $A_{1g}$\\
& & & &30 $A_{2g}$\\
& 3 & 3 & 400 &2$\times$66 $E_{2g}$\\
&   &   &    & 38 $B_{1u}$\\
& & & &30 $B_{2u}$\\
&  &  & &2$\times$66 $E_{1u}$\\\hline
& 2 & 4 & 225&\\
& 1 & 5 & 36 &$\vdots$\\
& 0 & 6 & 1&\\ \hline
\end{tabular}
\caption{Overview of the 6 particle states of benzene, sorted by
$S_z$ and symmetry. Orbitals with $A$- and $B$-type of symmetry
show no degeneracy, while $E$-type orbitals are doubly
degenerate.} \label{table:N6states}
\end{table}

Table \ref{table:N6states} shows an overview of the states of the
neutral molecule (the 6 particle states) sorted by $S_z$ and
symmetries. The eigenstates of the interacting benzene molecule
have either $A$-, $B$- or $E$-type symmetries. While orbitals
having $A$ or $B$ symmetries can only be spin degenerate, states
with an $E$ symmetry show an additional twofold orbital
degeneracy, essential for the explanation of the transport
features occurring in the meta configuration.

Transport at low bias is described in terms of transitions between ground
states with diferent particle number. Table \ref{table:GSsym} shows the
symmetries of the ground states (and of some first excited states) of
interacting benzene for all possible particle numbers. Ground state
transitions occur both between orbitally non-degenerate states
(with $A$ and $B$ symmetry), as well as between orbitally
degenerate and non-degenerate states ($E$- to $A$-type states).

The interacting benzene Hamiltonian commutes with all the symmetry
operations of the $D_{6h}$ point group, thus it has a set of common eigenvectors
with each operation. The element of $D_{6h}$
of special interest for the \emph{para} configuration is $\s_{\rm v}$,
\textit{i.e.}, the reflection about the plane through the contact
atoms and perpendicular to the molecular plane. The molecular
orbitals with $A$ and $B$ symmetry are eigenstates of $\sigma_{\rm
v}$ with eigenvalue $\pm1$, \textit{i.e.}, they are either
symmetric or antisymmetric with respect to the $\sigma_{\rm v}$
operation. The behavior of the $E-$type orbitals under $\s_{\rm
v}$ is basis dependent, yet one can always choose a basis in which
one orbital is symmetric and the other one antisymmetric.

Let us now consider the generic transition amplitude
$\bra{N}d^{\phd}_{\alpha\tau}\ket{N+1}$, where $d^{\phd}_{\alpha\tau}$ destroys
an electron of spin $\tau$ on the contact atom closest to the $\alpha$ lead. It
is useful to rewrite this amplitude in the form
\begin{eqnarray}
\bra{N}d^{\phd}_{\alpha\tau}\ket{N+1}= \bra{N} \sigma_{\rm
v}^\dagger\sigma^{\phd}_{\rm v} d^{\phd}_{\alpha\tau}\sigma_{\rm
v}^\dagger \sigma^{\phd}_{\rm v} \ket{N+1},\label{eq:NNp1trans}
\end{eqnarray}
where we have used the property $\sigma_{\rm
v}^\dagger\sigma^{\phd}_{\rm v}=1$. Since in the para
configuration both contact atoms lie in the mirror plane
$\sigma_{\rm v}$, it follows $\sigma^{\phd}_{\rm
v}d^{\phd}_\alpha\sigma_{\rm v}^\dagger=d^{\phd}_\alpha$. If the
participating states are both symmetric under $\sigma_{\rm v}$,
Eq.~\eqref{eq:NNp1trans}
becomes
\begin{eqnarray}
\bra{N,{\rm sym}} \sigma_{\rm v}^\dagger
d^{\phd}_{\alpha\tau}\sigma^{\phd}_{\rm v}
\ket{N+1,{\rm sym}}=\hspace{1cm}\nonumber\\
=\bra{N,{\rm sym}}d^{\phd}_{\alpha\tau}\ket{N+1,{\rm sym}}.
\end{eqnarray}
and analogously in the case that both states are antisymmetric.
For states with different symmetry it is
\begin{eqnarray}
\bra{N,{\rm sym}}d^{\phd}_{\alpha\tau}\ket{N+1,{\rm
antisym}}=\hspace{3cm}\nonumber\\
=-\bra{N,{\rm sym}}d^{\phd}_{\alpha\tau}\ket{N+1,{\rm
antisym}}=0.\hspace{0.2cm}
\end{eqnarray}
In other terms, there is a selection rule that
forbids transitions between symmetric and antisymmetric states.
Further, since the ground state of the neutral molecule is
symmetric, for the transport calculations in the para
configuration we select the effective Hilbert space containing
only states symmetric with respect to $\sigma_{\rm v}$.
Correspondingly, when referring to the $N$ particle ground state
we mean the energetically lowest \textit{symmetric} state. For
example in the case of 4 and 8 particle states it is the first
excited state to be the \textit{effective} ground state. In the
para configuration also the orbital degeneracy of the $E-$type
states is effectively cancelled due to the selection of the
symmetric orbital (see Table \ref{table:GSsym}).

Small violations of this selection rule, due \textit{e.g.} to
molecular vibrations or coupling to an electromagnetic bath,
result in the weak connection of different metastable electronic
subspaces. We suggest this mechanism as a possible explanation for
the switching and hysteretic behaviour reported in various
molecular junctions. This effect is not addressed in this work.

For a simpler analysis of the different transport characteristics
it is useful to introduce a unified geometrical description of the
two configurations. In both cases, one lead is rotated by an angle
$\phi$ with respect to the position of the other lead. Hence we
can write the creator of an electron in the right contact atom
$d^\dagger_{\rm R\tau}$ in terms of the creation operator of the
left contact atom and the rotation operator:
\begin{eqnarray}
d^\dagger_{\rm R\tau}=\mathcal{R}^\dagger_\phi d^\dagger_{\rm
L\tau}\mathcal{R}^{\phd}_\phi, \label{eq:d_R}
\end{eqnarray}
where $\mathcal{R}_\phi$ is the rotation operator for the
anticlockwise rotation of an angle $\phi$ around the axis
perpendicular to the molecular plane and piercing the center of
the benzene ring; $\phi=\pi$ for the para and $\phi=(2\pi/3)$ for
the meta configuration.

\begin{table}[h!]
\centering
\begin{tabular}{ccccc}
$N$ &\mbox{\hspace{0.1cm}}Degeneracy\mbox{\hspace{0.1cm}}
&\mbox{\hspace{0.1cm}}Energy[$e$V]\mbox{\hspace{0.1cm}} & Symmetry
& Symmetry behavior\\
& &(at $\xi=0$)& & under $\sigma_{\rm{v}}$\\
\hline
0  & 1 & 0 & $A_{1g}$ & sym\\  %\hline
1  & 2 & -22 & $A_{2u}$ & sym\\ % \hline
%& 4 & -19.50 & $E_{1g}$ & 2 sym, 2 asym \\ \hline
2 & 1 & -42.25 & $A_{1g}$ & sym\\%  \hline
%& 6 & -40.89 & $E_{1u}$ & 3 sym, 3 asym \\ \hline
3 & 4 & -57.42 & $E_{1g}$ & 2 sym, [2 antisym] \\ % \hline
%& 4 & -56.97 & $A_{1u}$ & asym \\ \hline
4 &  [3] & [-68.87] & [$A_{2g}$] & [antisym]\\  %\hline
& 2 & -68.37 & $E_{2g}$ & 1 sym, [1 antisym] \\ %\hline
5  & 4 & -76.675 & $E_{1g}$ & 2 sym, [2 antisym] \\ % \hline
%& 2 & -73.94 & $A_{2u}$ & sym \\ \hline
6 & 1 & -81.725 & $A_{1g}$ & sym\\   %\hline
%& 1 & -79.66 & $B_{1u}$ & sym\\ \hline
7 & 4 & -76.675 & $E_{2u}$ & 2 sym, [2 antisym]\\  %\hline
%& 2 & -73.94 & $B_{2g}$ & sym\\ \hline
8 & [3] & [-68.87] & [$A_{2g}$] & [antisym]\\ % \hline
& 2 & -68.37 & $E_{2g}$ & 1 sym, [1 antisym] \\ %\hline
9 & 4 & -57.42 &$E_{2u}$ &  2 sym, [2 antisym]\\ % \hline
%& 4 & -57.97 & $B_{1g}$ & asym \\ \hline
10 & 1 & -42.25 & $A_{1g}$ & sym\\%  \hline
%& 6 & -40.89 & $E_{1u}$ & 3 sym, 3 asym \\ \hline
11 & 2 & -22 & $B_{2g}$ & sym\\  %\hline
%& 4 & -19.50 & $E_{2u}$ & 2 sym, 2 asym \\ \hline
12 & 1 & 0 & $A_{1g}$ & sym
\end{tabular}
\caption{Degeneracy, energy and symmetry of the ground states of
the isolated benzene molecule for different particle numbers. We
choose the on-site and inter-site Coulomb interactions to be
$U=10\,e{\rm V}$, $V=6\,e{\rm V}$, and the hopping to be
$b=-2.5\,e{\rm V}$. Notice, however, that screening effects from
the leads and the dielectric are expected to renormalize the
energy of the benzene many-body states.} \label{table:GSsym}
\end{table}

The energy eigenstates of the interacting Hamiltonian of benzene can be
classified also in terms their quasi-angular momentum.
In particular, the eigenstates of the $z$-projection of the quasi angular
momentum are the ones that diagonalize all operators $\mathcal{R}_\phi$ the
with angles multiples of $\pi/3$. The corresponding eigenvalues are phase
factors $e^{-i\ell\phi}$ where $\hbar\ell$, the quasi-angular momentum of the
state, is an integer multiple of $\hbar$. The discrete rotation operator of
an angle $\phi=\pi$ ($C_2$ symmetry operation), is the one relevant for the para
configuration. All orbitals are eigenstates of the $C_2$ rotation with the
eigenvalue $\pm 1$.

The relevant rotation operator for the meta configuration correspond to an angle
$\phi=2\pi/3$ ($C_3$ symmetry operation). Orbitals with an $A$ or $B$ symmetry
are eigenstates of
this operator with the eigenvalue $+1$ (angular momentum $\ell = 0$ or $\ell
=
3$). Hence we can already predict that there will be no difference based on
rotational symmetry between the para and the meta configuration for
transitions between states involving $A$- and $B$-type symmetries.
Orbitals with $E$ symmetry however behave quite differently under
the $C_3$ operation. They are the pairs of states of angular momenta $\ell =
\pm 1$ or $\ell = \pm 2$.
The diagonal form of the rotation operator on the two-fold degenerate subspace
of $E$-symmetry reads:
\begin{eqnarray}
C_3 =
\left(\begin{matrix}
e^{-|\ell|\cdot\frac{2\pi}{3}i} & 0\\
0 & e^{|\ell|\cdot\frac{2\pi}{3}i}
\end{matrix}\right)
\end{eqnarray}
For the two-fold orbitally degenerate 7-particle ground states $|\ell|=2$. This
analysis in terms of the quasi-angular momentum makes easier the calculation
of the fundamental interference condition \eqref{eq:B_interf_condition} given in
the introduction. In fact the following relation holds between the transition
amplitudes of the 6 and 7 particle ground states:

\begin{equation}
\begin{split}
\gamma_{\ell R} &\equiv \bra{ 7_g\ell \tau} d^\dagger_{R\tau}\ket{6_g} \\
               &=      \bra{ 7_g\ell \tau}
                       \mathcal{R}^\dagger_\phi d^\dagger_{\rm
L\tau}\mathcal{R}^{\phd}_\phi,
                       \ket{6_g} = e^{-i\ell\phi}\gamma_{\ell L}
\end{split}
\end{equation}
and \eqref{eq:B_interf_condition} follows directly.
%///////////////////////////////////////////////////////////////////////////////

%///////////////////////////////////////////////////////////////////////////////
\section{Transport calculations: Fully symmetric setup}\label{clean}
%///////////////////////////////////////////////////////////////////////////////

\begin{figure}[h!]
  \includegraphics[width=1.0\columnwidth,angle=0]{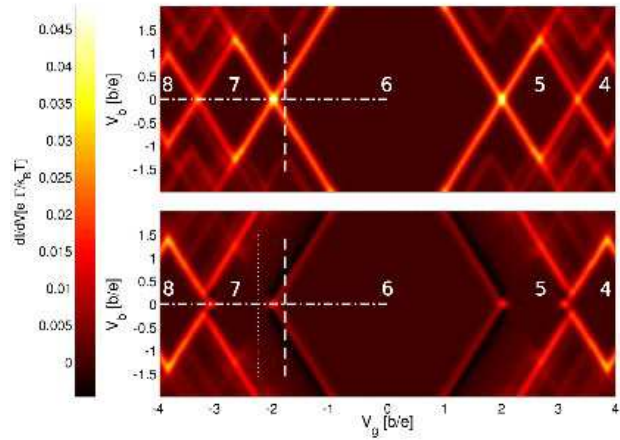}
  \caption{
  (color online) Stability diagram for the benzene I-SET
  contacted in the para (above) and meta (below) configuration.
  Dot-dashed lines highlight the conductance cuts presented
  in Fig.~\ref{fig3}, the dashed lines the regions corresponding
  to the current traces presented in Fig.~\ref{fig4} and Fig.~\ref{fig6}, the
  dotted line the region corresponding to the current trace presented in
  Fig.~\ref{fig5}. The parameters used are $U=4|b|,\,V = 2.4|b|,
  \,k_{\rm B}T = 0.04|b|,\,\hbar\Gamma_{\rm L} = \hbar\Gamma_{\rm R} =
10^{-3}|b|$.
  }
  \label{fig2}
\end{figure}

With the knowledge of the eigenstates and eigenvalues of the
Hamiltonian for the isolated molecule, we implement
Eq.~\eqref{eq:GME} and look for a stationary solution. The
symmetries of the eigenstates are reflected in the transition
amplitudes contained in the GME. We find numerically its
stationary solution and calculate the current and the differential
conductance of the device. In Fig.~\ref{fig2} we present the
stability diagram for the benzene I-SET contacted in the para
(upper panel) and meta position (lower panel). Bright ground state
transition lines delimit diamonds of zero differential conductance
typical for the Coulomb blockade regime, while a rich pattern of
satellite lines represents the transitions between excited states.
Though several differences can be noticed, most striking are the
suppression of the linear conductance, the appearance of negative
differential conductance (NDC) and the strong suppression of the current at the
right(left) border of the 7 (5) particle diamond when passing from
the para to the meta configuration. All these features are different
manifestations of the interference between orbitally degenerate
states and ultimately reveal the specific symmetry of benzene.

%-------------------------------------------------------------------------------
\subsection{Linear conductance}\label{LinCond_clean}
%-------------------------------------------------------------------------------
We study the linear transport regime both numerically and
analytically. For the analytical calculation of the conductance we
consider the low temperature limit where only ground states with
$N$ and $N+1$ particles have considerable occupation
probabilities, with $N$ fixed by the gate voltage. Therefore only
transitions between these states are relevant and we can treat
just the terms of (\ref{eq:GME}) with $N$ and $N+1$ particles and
the ground state energies $E_{\rm g,N}$ and $E_{\rm g,N+1}$,
respectively. A closer look to (\ref{eq:GME}) reveals that the
spin coherences are decoupled from the other elements of the
density matrix. Thus we can set them to zero, and write
(\ref{eq:GME}) in a block diagonal form in the basis of the
ground states of N and N+1 particles. Additionally, since the total Hamiltonian
$H$ is symmetric
in spin, the blocks of the GME with the same particle but
different spin quantum number $\t$ must be identical. Finally,
since around the resonance the only populated states are the $N$
and $N+1$ particle states, the conservation of probability implies
that:
\begin{equation}
 1=\sum_{n}\s^{\rm N}_{nn}+\sum_{m}\s^{\rm N+1}_{mm},
\end{equation}
where $\s^{\rm N}_{nn}$ is the population of the N-particle ground state and
$n$ contains the orbital and spin quantum numbers.
With all these observations we can reduce \eqref{eq:GME} to a much
smaller set of coupled differential equations, that can be treated
analytically. The stationary solution of this set of equations can
be derived more easily by neglecting the energy non-conserving
terms in \eqref{eq:GME}. These are contained in the elements of
the GME describing the dynamics of the coherences between
orbitally degenerate states. With this simplification we derive an
analytical formula for the conductance close to the resonance
between $N$ and $N+1$ particle states as the first order
coefficient of the Taylor series of the current in the bias:
\begin{equation}
\begin{split}
G_{\rm N,N+1}(\Delta E) =&
 2e^2\frac{\Gamma_{\rm L} \Gamma_{\rm R}}{\Gamma_{\rm L} + \Gamma_{\rm R}}
 \Lambda_{\rm N,N+1} \times\\
 &\times\left[
 -\frac{S_{\rm N}S_{\rm N+1} f'(\Delta E)}
       {(S_{\rm N+1}-S_{\rm N})f(\Delta E)+S_{\rm N}}
 \right]
\label{eq:G_NNp1}
\end{split}
\end{equation}
where $\Delta E = E_{\rm g,N} - E_{\rm g,N+1} + eV_{\rm g}$ is the
energy difference between the benzene ground states with $N$ and
$N+1$ electrons diminished by a term linear in the gate voltage.
Interference effects are contained in the overlap factor
$\Lambda_{\rm N,N+1}$:
\begin{eqnarray}
\Lambda_{\rm N,N+1} = \frac{
    \Big|
        \sum \limits_{nm\tau}
        \langle N,n|d^{\phd}_{\rm L\tau}|N\!+\!1,m\rangle
        \!
        \langle N\!+\!1,m |d^{\dagger}_{\rm R\tau}|N,n\rangle
    \Big|^2
 }
 {
    S_{\rm N}S_{\rm N+1}
    \sum \limits_{nm\alpha\tau}
    \Big|
        \langle
        N,n|d^{\phd}_{\alpha\tau}|N\!+\!1,m
        \rangle
    \Big|^2
 },
\nonumber \label{eq:urlambda}
\end{eqnarray}
where $n$ and $m$ label the $S_{\rm N}$-fold and $S_{\rm
N+1}$-fold degenerate ground states with $N$ and $N+1$ particles,
respectively. In order to make the interference effects more
visible we remind that $d^{\dagger}_{\rm R\tau} =
\mathcal{R}^{\dagger}_\phi d^{\dagger}_{\rm L\tau}
\mathcal{R}^{\phd}_\phi$, with $\phi = \pi$ for the para while
$\phi = 2\pi/3$ for the meta configuration. Due to the behaviour
of all eigenstates of $H^0_{\rm ben}$ under discrete rotation
operators with angles multiples of $\pi/3$, we can rewrite the
overlap factor:
\begin{eqnarray}
\Lambda_{\rm N,N+1} = \frac{
    \Big|
        \sum \limits_{nm\tau}
        |\langle N,n|d^{\phd}_{\rm L\tau}|N\!+\!1,m\rangle|^2
        e^{i\phi_{nm}}
    \Big|^2}
{S_{\rm N}S_{\rm N+1}
    2\sum \limits_{nm\tau}
    \Big|\langle
    N,n|d^{\phd}_{\rm L\tau}|N\!+\!1,m
    \rangle\Big|^2}, \label{eq:lambda}
\end{eqnarray}
where $\phi_{nm}$ encloses the phase factors coming from the
rotation of the states $\ket{N,n}$ and $\ket{N+1,m}$.

The energy non-conserving terms neglected in \eqref{eq:G_NNp1}
influence only the dynamics of the coherences between orbitally
degenerate states. Thus, Eq.~\eqref{eq:G_NNp1} provides an exact
description of transport for the para configuration, where orbital
degeneracy is cancelled. Even if Eq.~\eqref{eq:G_NNp1}
captures the essential mechanism responsible for the conductance suppression, we
have derived an exact analytical formula also for the meta
configuration and we present it in Appendix~\ref{appendixPP}.

In Fig.~\ref{fig3} we present an overview of the results of both
the para and the meta configuration. A direct comparison of the
conductance (including energy non-conserving terms) in the two
configurations is displayed in the upper panel. The lower panel
illustrates the effect of the energy non-conserving terms on the
conductance in the meta configuration. The number of $p_z$
electrons on the molecule and the symmetry of the lowest energy
states corresponding to the conductance valleys are reported. The
symmetries displayed in the upper panel belong to the (effective)
ground states in the para configuration, the corresponding
symmetries for the meta configuration are shown in the lower
panel.

Fig.~\ref{fig3} shows that the results for the para and the meta
configuration coincide for the $10 \leftrightarrow 11$ and $11
\leftrightarrow 12$ transitions. The ground states with
$N=10,11,12$ particles have $A-$ or $B-$type symmetries, they are
therefore orbitally non-degenerate, no interference can occur and
thus the transitions are invariant under configuration change.
For every other transition we see a noticeable difference between
the results of the two configurations (Fig.~\ref{fig3}). In all
these transitions one of the participating states is orbitally
degenerate. First we notice that the linear conductance peaks for
the $7 \leftrightarrow 8$ and $8 \leftrightarrow 9$ transitions in
the para configuration are \emph{shifted} with respect to the
corresponding peaks in the meta configuration. The selection of an
effective symmetric Hilbert space associated to the para
configuration reduces the total degeneracy by cancelling the
orbital degeneracy. In addition, the ground state energy of the 4
and 8 particle states is different in the two configurations,
since in the para configuration the \emph{effective} ground state
is in reality the first excited state. The degeneracies $S_{\rm
N}, S_{\rm N+1}$ of the participating states as well as the ground
state energy are both entering the degeneracy term of
Eq.~\eqref{eq:G_NNp1}
\begin{equation}
\Delta=-\frac{f'(\Delta E)}{(S_{\rm N+1}-S_{\rm N})f(\Delta
E)+S_{\rm N}},
\end{equation}
and determine the shift of the conductance peaks.\\
\begin{figure}[h!]
  \includegraphics[width=1\columnwidth,angle=0]{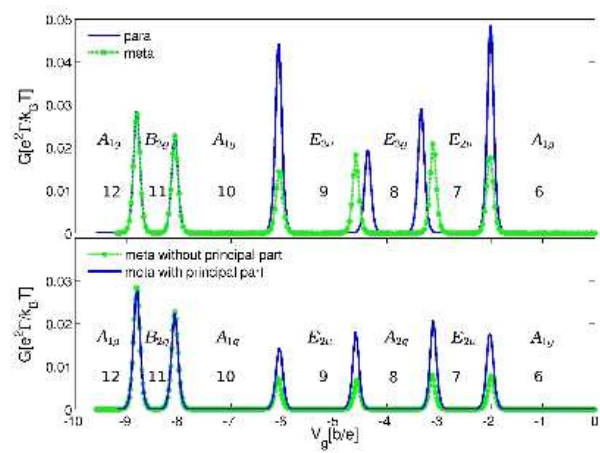}
  \caption{(color online) Conductance of the benzene I-SET as a function of the
gate voltage.
  Clearly visible are the peaks corresponding to the transitions between ground
states with
  $N$ and $N+1$ particles. In the low conductance valleys the state of the
system
  has a definite number of particles and symmetry as indicated in the upper
panel for the para,
  in the lower for the meta configuration. Selective conductance  suppression
  when changing from the meta to the para configuration is observed.}
  \label{fig3}
\vspace{-2mm}
\end{figure}

Yet, the most striking effect regarding transitions with orbitally
degenerate states participating is the \emph{systematic
suppression} of the linear conductance when changing from the para
to the meta configuration. The suppression is appreciable despite
the conductance enhancement due to the energy non-conserving terms
(see Fig.~\ref{fig3}, lower panel). Thus, we will for simplicity
discard them in the following discussion.

The conductance suppression is determined by the
combination of two effects: the reduction to the symmetric Hilbert
space in the para configuration and the interference effects
between degenerate orbitals in the meta configuration. The reduction to the
symmetric Hilbert space implies also a lower number of conducting
channels (see Table \ref{table:CompParaMeta}). One would expect a
suppression of transport in the para configuration. As we can see
from Table \ref{table:CompParaMeta} on the example of the $6
\leftrightarrow 7$ transition peak, $\Delta_{\rm max}$ is higher
in the para configuration but not enough to fully explain the
difference between the two configurations.
\begin{table}[h!]
\centering
\begin{tabular}{c|c|c|c}
& \# Channels & Overlap factor & Degeneracy term \\
& $S_{\rm N} S_{N+1}$& $\Lambda$ & $\Delta_{\rm max}$ [1/$k_{\rm
B}T$]\\\hline
PARA & 2 & $C$ & 0,17\\
META & 4 & $\frac{1}{8}C$ & 0,11
\end{tabular}
\caption{Number of channels participating to transport, overlap
factor and resonance value of the degeneracy term in the para and
the meta configuration for the $6 \leftrightarrow 7$ transition
peak. It is $C=|\bra{6_{\rm g}}d_{\rm L\t}\ket{7_{\rm g}\l\t}|^2
$, where $\t$ and $\ell$ are the spin and the quasi angular momentum quantum
numbers, respectively. The values of $\Delta_{\rm max}$ are given
for $k_{\rm B}T = 0.04|b|$}. \label{table:CompParaMeta}.
\end{table}

The second effect determining transport is the interference
between the $E$-type states, which is accounted for in the overlap
factor $\Lambda$. The overlap factor is basis independent, thus we can
write the transition probabilities for the $6 \leftrightarrow 7$
transition as $|\langle 6_{\rm g}|d^{\phd}_{\rm L\tau}|7_{\rm
g}\,\ell\,\t\rangle|^2=C$, where $\t$ and $\ell$ are the spin and
the quasi-angular momentum quantum number, respectively.
The transition probabilities have the same value, since all four 7
particle states are in this basis equivalent (see
Appendix~\ref{appendixC}). Under the $C_2$ rotation the symmetric
7 particle ground state does not acquire any phase factor. Under
the $C_3$ rotation however, the two orbitally degenerate states
acquire different phase factors, namely $e^{\frac{4\pi}{3}i}$ and
$e^{-\frac{4\pi}{3}i}$, respectively. Thus the overlap factors
$\Lambda$ for the $6 \leftrightarrow 7$ transition are:
\begin{eqnarray}
\Lambda_{\rm para} &=&
        \frac{1}{2\cdot 8C}\cdot\left|4C\right|^2=C,\nonumber\\
\Lambda_{\rm meta} &=&
        \frac{1}{4\cdot8C}\cdot\left|2C e^{+\frac{4\pi}{3}i}+
        2C e^{-\frac{4\pi}{3}i}\right|^2=\frac{1}{8}C\nonumber.
\end{eqnarray}
The linear conductance is determined by the product between the
number of conducting channels, the overlap factor and the
degeneracy term. Yet, it is the \emph{destructive interference}
between degenerate $E$-type orbitals, accounted for in the overlap
factor $\Lambda$, that gives the major contribution to the strong
suppression of the conductance in the meta configuration.
%-------------------------------------------------------------------------------

%-------------------------------------------------------------------------------
\subsection{Negative differential conductance (NDC) and current
blocking}\label{NDC_clean}
%-------------------------------------------------------------------------------

\begin{figure}[h!]
  \includegraphics[width=\columnwidth,angle=0]{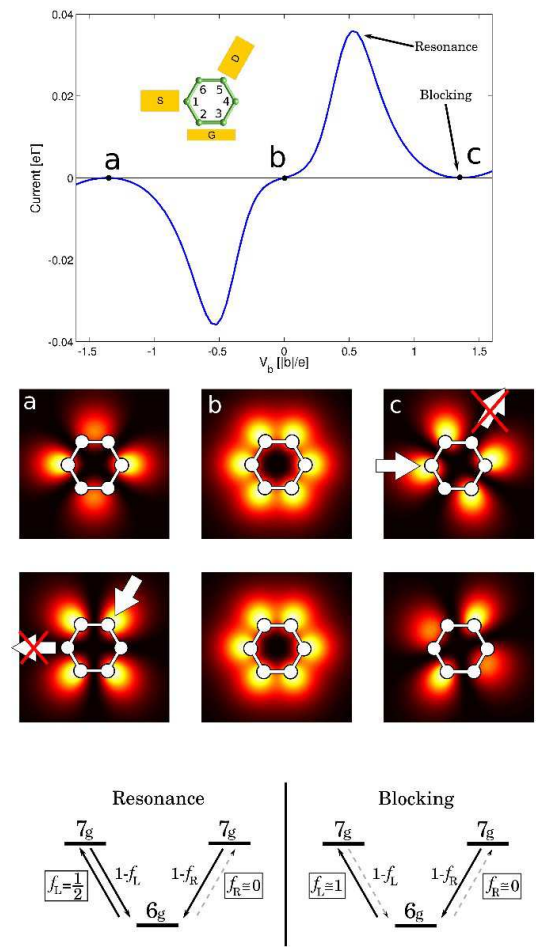}
  \caption{(color online) Upper panel - Current through the benzene I-SET
  in the meta configuration calculated at bias and gate voltage conditions
indicated by the dashed line of Fig. \ref{fig2}. A pronounced NDC with current
blocking is visible.
  Middle panels - Transition probabilities between the 6 particle and each of
the two 7 particle ground states for bias voltage values labelled $a-c$ in the
upper
panel. The transition to a blocking state is visible in the upper (lower)
  part of the $c$ ($a$) panels. Lower panels - Sketch of the energetics for the
$6\rightarrow7$
transition in the meta configuration at bias voltages
corresponding to the resonance current peak and current blocking as indicated in
the upper panel of this figure.}
  \label{fig4}
\end{figure}

Interference effects between orbitally degenerate states are also
affecting non-linear transport and producing in the \emph{meta}
configuration current blocking and thus NDC at the border of the 6
particle state diamond (Fig. \ref{fig2}). The upper panel of Fig.
\ref{fig4} shows the current through the benzene I-SET contacted
in the meta configuration as a function of the bias voltage. The
current is given for parameters corresponding to the white dashed
line of Fig.~\ref{fig2}. In this region only the 6 and 7 particle
ground states are populated.

At low bias the 6 particle state is mainly occupied. As the bias
is raised, transitions $6 \leftrightarrow 7$ occur and current
flows. Above a certain bias threshold a blocking state is
populated and the current drops. For the understanding of this
non-linear current characteristics, we have to take into account
energy conservation, the Pauli exclusion principle and the
interference between participating states. For the visualization
of the interference effects, we introduce the transition
probability (averaged over the $z$ coordinate and the spin
$\sigma$):
\begin{equation}
P(x,y;n,\,\t)=\lim_{L \to \infty} \sum_\s\frac{1}{2L}
\int_{-L/2}^{L/2}\hspace{-0.1cm}{\rm d}z |\langle 7_{\rm
g}\,n\,\t| \psi^{\dagger}_\s({\bf r})| 6_{\rm g}\rangle|^2
\label{eq:transition_surf}
\end{equation}
for the \emph{physical} 7 particle basis, \textit{i.e.}, the 7
particle basis that diagonalizes the stationary density matrix at
a fixed bias. Here $\t$ is the spin quantum number, $n=1,2$ labels
the two states of the physical basis which are linear combinations
of the orbitally degenerate states $\ket{7_{\rm g}\ell\t}$ and can be
interpreted as conduction channels.
Each of the central panels of Fig. \ref{fig4} are surface plots of
\eqref{eq:transition_surf} at the different bias voltages {\it a}-{\it c}.
The 7 particle ground states can interfere and thus generate nodes
in the transition probability at the contact atom close to one or
the other lead, but, in the meta configuration, never at both
contact atoms at the same time.

Energetic considerations are illustrated in the lower panels of Fig.~\ref{fig4}
for two key points of the current curve at positive biases. The left panel
corresponds to the resonance peak of the current. Due to energy conservation,
electrons can enter the molecule only from the left lead. On the contrary the
exit is allowed at both leads. The current is suppressed when transitions
occur to a state which cannot be depopulated (a blocking state).
Since, energetically, transmissions to the 6 particle state are allowed at both
leads, each 7 particle state can always be depopulated and no blocking
occurs.

The current blocking scenario is depicted in the lower right panel of
Fig.~\ref{fig4}. For large positive bias the transition from a 7
particle ground state to the 6 particle ground state is
energetically forbidden at the left lead. Thus, for example, the $c$ panel in
Fig.~\ref{fig4} visualizes the current blocking situation yielding NDC: while
for both channels there is a non-vanishing transition probability from the
source lead to the molecule, for the upper channel a node prevents an electron
from exiting to the drain lead. In the long time limit the
blocking state gets fully populated while the non-blocking state
is empty. At large negative bias the blocking scenario is depicted
in the panel $a$ that shows the left-right symmetry obtained by a reflection
through a plane perpendicular to the molecule and passing through the carbon
atoms atoms 6 and 3.\\
We remark that only a description that retains coherences between
the degenerate 7 particle ground states correctly captures NDC at
both positive and negative bias.

\begin{figure}[h!]
\includegraphics[width=\columnwidth,angle=0]{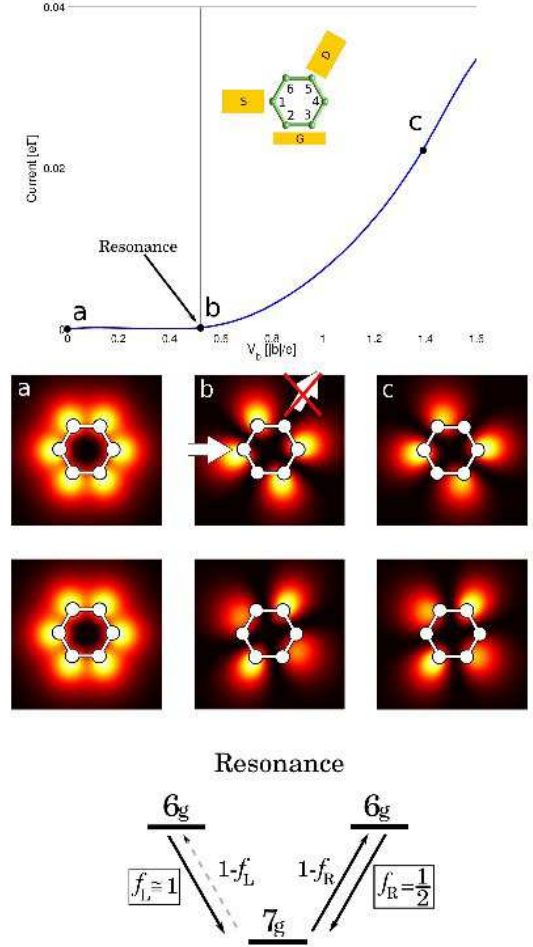}
\caption{(color online) Upper panel - Current through the benzene
I-SET in the meta configuration calculated at  bias and gate
voltage conditions indicated by the dotted line of Fig.
\ref{fig2}. No NDC is visible. Middle panels - Transition
probabilities between each of the 7 particle and the 6 particle
ground state for bias voltage values labelled $a-c$ in the upper
panel. Lower panel - Sketch of the energetics for the $7\rightarrow6$
transition in the meta configuration at bias voltage corresponding
to the expected resonance peak. (compare to Fig.~\ref{fig4}).} \label{fig5}
\end{figure}

In contrast to the $6 \rightarrow 7$ transition, one does \emph{not}
observe NDC at the border of the 7 particle Coulomb diamond, but
rather a strong suppression of the current. The upper panel of
Fig.~\ref{fig5} shows the current through the benzene I-SET
contacted in the meta configuration as a function of the bias
voltage corresponding to the white dotted line of Fig.~\ref{fig2}.
The middle panels show the transition probabilities between each of
the 7 particle and the 6 particle ground state.

The lower panel of Fig.~\ref{fig5} shows a sketch of the energetics at positive
bias
corresponding to the ``expected'' resonance peak. Here electrons
can enter the molecular dot at both leads, while the exit is
energetically forbidden at the left lead. Thus, if the system is
in the 7 particle state which is blocking the right lead, this
state cannot be depopulated, becoming the blocking state.\\
On the other hand, transitions from the 6 particle ground state to
both 7 particle ground states are equally probable. Thus the
blocking state will surely be populated at some time. The upper
plot of the $b$ panel in Fig.~\ref{fig5} shows the transition
probability to the blocking state that accepts electrons from the
source lead but cannot release electrons to the drain.\\
As just proved, in this case the current blocking situation occurs
already at the resonance bias voltage. For a higher positive bias,
the transition probability from the blocking state at the drain
lead increases and current can flow. This effect, though, can be
captured only by taking into account also the energy
non-conserving terms in \eqref{eq:GME}.

\begin{figure}[h!]
  \includegraphics[width=0.9\columnwidth,angle=0]{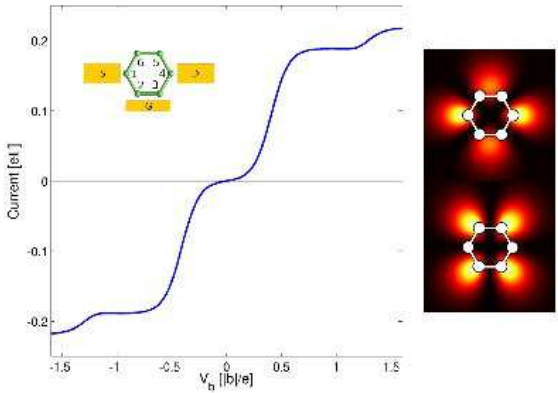}
  \caption{(color online) Left panel - Current through the benzene I-SET
  in the para configuration calculated at  bias and gate voltage conditions
indicated by   the dashed line of Fig.~\ref{fig2}. No interference effects are
visible. Right panels - Transition probabilities between the 6 particle and the
symmetric   and antisymmetric 7 particle ground states.}
  \label{fig6}
\end{figure}

In the \emph{para} configuration, the current as a function of the
bias voltage is shown in Fig.~\ref{fig6}. The current is given for parameters
corresponding to the white dashed line of Fig.~\ref{fig2}. In this case, no
interference
effects are visible. We see instead the typical step-like behavior
of the current in the Coulomb blockade regime.\\
The panels on the right are the surface plots of
\begin{equation}
P(x,y; \t)=\lim_{L \to \infty} \sum_\s\hspace{-0.1cm}\frac{1}{2L}
\hspace{-0.1cm}\int_{-L/2}^{L/2}\hspace{-0.15cm}{\rm d}z |\langle
7_{\rm g}\,\t;{\rm (a)sym}| \psi^{\dagger}_\s({\bf r})| 6_{\rm
g}\rangle|^2.
\end{equation}
The upper plot shows the transition probability to the symmetric 7
particle state, the lower to the antisymmetric. Remember that in
the para configuration only the symmetric states contribute to
transport. Evidently the symmetric state is in the para
configuration non-blocking. Additionally, since the coherences
between orbitally degenerate states and therefore the energy
non-conserving terms do not play any role in the transport, the
physical basis states are not bias dependent. Thus in the para
configuration there are always non-blocking states populated and
no NDC can occur.
%///////////////////////////////////////////////////////////////////////////////

%///////////////////////////////////////////////////////////////////////////////
\section{Reduced symmetry}\label{ReducedSym}
%///////////////////////////////////////////////////////////////////////////////

In this section we study the effect of reduced symmetry on the
results presented previously. We generalize the model Hamiltonian by taking into
account the perturbations on the molecule due to
the contacts and the bias voltage. The contact between molecule
and leads is provided by different anchor groups. These linkers are
coupled to the contact carbon atoms over a $\s$ bond thus
replacing the corresponding benzene hydrogen atoms. Due to the
orthogonality of $\p$ and $\s$ orbitals, the anchor groups affect
in first approximation only the $\s$ orbitals of benzene. In
particular the different electron affinity of the atoms in the
linkers imply a redistribution of the density of $\s$ electrons.
Assuming that transport is carried by $\pi$ electrons only, we
model the effect of this redistribution as a change in the on-site
energy for the $p_z$ orbitals of the contact carbon atoms:
\begin{eqnarray}
H^\prime_{\rm ben}:=H_{\rm contact}=\xi_{\rm c}\sum_{\alpha\s}
d^{\dagger}_{\alpha\s}d^{\phd}_{\alpha\s},\hspace{0.3cm}\alpha=L,R
\label{eq:H_cont}
\end{eqnarray}
where $R=4,5$, respectively, in the para and meta configuration,
$L=1$ in both setups.

We also study the effect of an external bias on the benzene I-SET.
In particular we release the strict condition of potential drop
all concentrated at the lead-molecule interface. Nevertheless, due
to the weak coupling of the molecule to the leads, we assume that
only a fraction of the bias potential drops across the molecule.
For this residual potential we take the linear approximation
$V_{\rm b}(\textbf{r})=-\frac{V_{\rm b}}{a}(\textbf{r}\cdot
\hat{\textbf{r}}_{\rm sd}/a_0)$, where we choose the center of the
molecule as the origin and $\hat{\textbf{r}}_{\rm sd}$ is the
unity vector directed along the source to drain direction.
$a_0=1.43\,{\rm \AA{}}$ is the bond length between two carbon
atoms in benzene, $a$ is the coefficient determining the intensity
of the potential drop over the molecule. Since the $p_z$ orbitals
are strongly localized, we can assume that this potential will not
affect the inter-site hopping, but only the on-site term of the
Hamiltonian:
\begin{eqnarray}
H^\prime_{\rm ben}:=H_{\rm bias}=e\sum_{i\s}\xi_{{\rm
b}_i}d^{\dagger}_{i\s}d^{\phd}_{i\s} \label{eq:H_bias}
\end{eqnarray}\\
 with $\xi_{{\rm b}_i}=\int d{\bf r}
 \hspace{0.1cm}p_{z}({\bf r}- {\bf R}_i)
 V_{\rm b}({\bf r})p_{z}({\bf r}-{\bf R}_i)$.\\

Under the influence of the contacts or the bias potential, the
symmetry of the molecule changes. Table~\ref{table:RedSym} shows
the point groups to which the molecule belongs in the perturbed
setup. This point groups have only $A$- and $B$-type reducible
representations. Thus the corresponding molecular orbitals do not
exhibit orbital degeneracy.

No interference effects influence the transport in the para configuration. Thus
we do not expect its transport characteristics to be qualitatively modified by
the new set up with the corresponding loss of degeneracies.\\

In the meta configuration on the other hand, interferences between
orbitally degenerate states play a crucial role in the explanation
of the occurring transport features. Na\"{\i}vely one would
therefore expect that neither conductance suppression nor NDC and
current blocking occur in a benzene I-SET with reduced symmetry.
\begin{table}[h!]
\centering
\begin{tabular}{c|cc}
& \hspace{0.3cm}PARA\hspace{0.3cm} &
\hspace{0.3cm}META\hspace{0.3cm} \\\hline
Contact perturb. & $D_{2h}$ & $C_{2v}$\\
Bias perturb. & $C_{2v}$ & $C_{2v}$
\end{tabular}
\caption{Point groups to which the molecule belongs under the
influence of the contacts and the external bias potential.}
\label{table:RedSym}
\end{table}
Yet we find that, under certain conditions, the mentioned
transport features are robust under the lowered symmetry.

The perturbations due to the contacts and the bias lead to an
expected level splitting of the former orbitally degenerate
states.  Very different
current-voltage characteristics are obtained depending of the relation between
the energy splitting $\delta E$ and other two important energy scales of the
system: the tunnelling rate $\Gamma$ and the temperature $T$. In particular,
when $·\delta E \ll \Gamma \ll T$, interference phenomena persist. In contrast
when $\Gamma < \delta E \ll T$ interference phenomena disappear, despite the
fact that, due to temperature broadening, the two states still can not be
resolved. In this regime, due to the asymmetry in the tunnelling rates
introduced by the perturbation, standard NDC phenomena, see Fig.~\ref{fig8},
occur.

In the absence of perfect degeneracy, we abandon the strict secular
approximation scheme that would discard the coherences in the density matrix
between states with different energies. We adopt instead a softer approximation
by retaining also coherences between quasi-degenerate states. Since they have
Bohr frequencies comparable to the tunnelling rate, they influence the
stationary density matrix. Formulas for the GME and the current taking into
account these
coherence terms are presented in Appendix~\ref{appendixGME}.
\begin{figure}[h!]
  \includegraphics[width=0.9\columnwidth,angle=0]{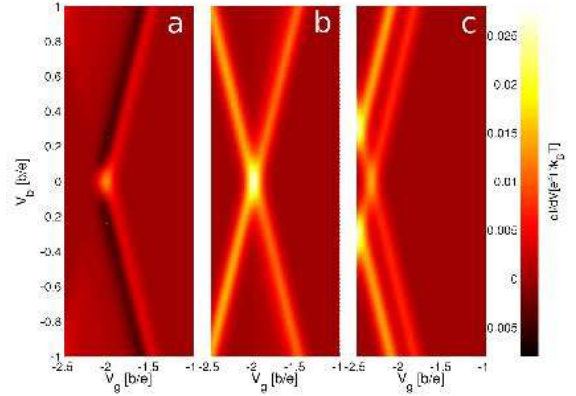}
\caption{(color online) Closeup views of the stability diagram
around the $6\leftrightarrow7$ resonance for the system under
contact perturbation. The perturbation strength grows from left to
right The parameter that describes the contact effect assumes the values $\xi_c
= 0.15\Gamma,\,2\Gamma,\,15T$ from left to right respectively and $T = 10\Gamma$
.}
  \label{fig7}
\end{figure}

Fig.~\ref{fig7}  shows from left to right closeup views of the
stability diagram for the setup under the influence of increasing
\emph{contact perturbation} around the $6\leftrightarrow7$
resonance. The orbital degeneracy of the 7 particle states is
lifted and the transport behavior for the $6\leftrightarrow7$
transition depends on the energy difference between the formerly
degenerate 7 particle ground states. In panel $a$ the energy
difference is so small that the states are quasi-degenerate:
$\delta E \ll \hbar\G \ll k_{\rm B}T$. As expected, we recover NDC
at the border of the 6 particle diamond and current suppression at
the border of the 7 particle diamond, like in the unperturbed
setup.\\
Higher on-site energy-shifts correspond to a larger level
spacing. Panel $b$ displays the situation in which the latter is
of the order of the level broadening, but still smaller than the
thermal energy ($\delta E \simeq \hbar\G \ll k_{\rm B}T$): no
interference causing NDC and current blocking can occur. Yet, due
to thermal broadening, we cannot resolve the two 7 particle
states.\\
Eventually, panel $c$ presents the stability diagram for the case
$\delta E > k_{\rm B}T > \hbar\G$: the level spacing between the 7
particle ground and first excited state is now bigger than the
thermal energy, thus the two transition lines corresponding to
these states are clearly visible at the border of the 6 particle
stability diamond.

\begin{figure}[h!]
  \includegraphics[width=0.9\columnwidth,angle=0]{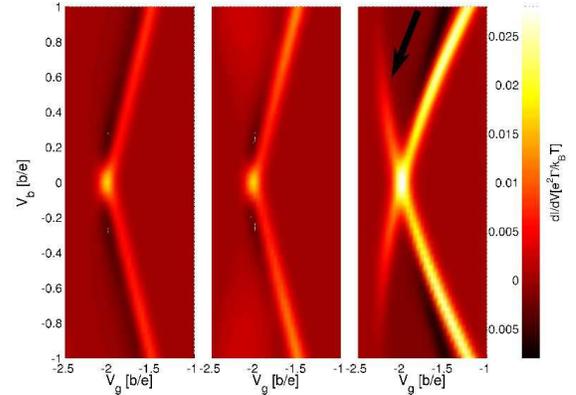}
\caption{(color online) Closeup views of the stability diagram
around the $6\leftrightarrow7$ resonance for the system under the
effect of the bias potential, displayed for different strengths of
the electrostatic potential drop over the molecule. The parameter that describe
the strength of the electrostatic drop overthe molecule assumes the values $a =
25,\,12,\,0.6$ from left to right respectively.}
  \label{fig8}
\end{figure}

Fig.~\ref{fig8} shows closeup views of the stability diagram for
the setup under the influence of the \emph{bias perturbation} at
the border of the 6 and 7 particle diamonds. The same region is
plotted for different strengths of the external potential over the
molecule. In contrast to the contact perturbation, the amount of
level splitting of the former degenerate states is here bias
dependent. This fact imposes a bias window of interference
visibility. The bias must be small enough, for the 7 particle
states to be quasi-degenerate and at the same time bigger than the
thermal energy, so that the occurring NDC is not obscured by the
thermally broadened conductance peak. A strong electrostatic
potential perturbation closes the bias window and no interference
effect can be detected.\\
Panel $a$ of Fig.~\ref{fig8} represents the weak perturbation
regime with no qualitative differences with the unperturbed case.
The typical fingerprints of interference (NDC at the border of the
6 particle diamond and current blocking for the $7\rightarrow6$
transition) are still visible for intermediate perturbation
strength (panel $b$) but this time only in a limited bias window.
Due to the perturbation strength, at some point in the bias, the
level splitting is so big that the quasi-degeneracy is lifted and
the interference effects destroyed. In panel $c$ the
quasi-degeneracy is lifted in the entire bias range. There is NDC
at the border of the 6 particle diamond, but is not accompanied by
current blocking as proved by the excitation line at the border of
the 7 particle diamond (see arrow): no interference occurs. The NDC is here
associated to the sudden opening of a slow current channel, the
one involving the 6 particle ground state and the 7 particle (non-degenerate)
excited state (standard NDC).

\begin{figure}[h!]
  \includegraphics[width=0.9\columnwidth,angle=0]{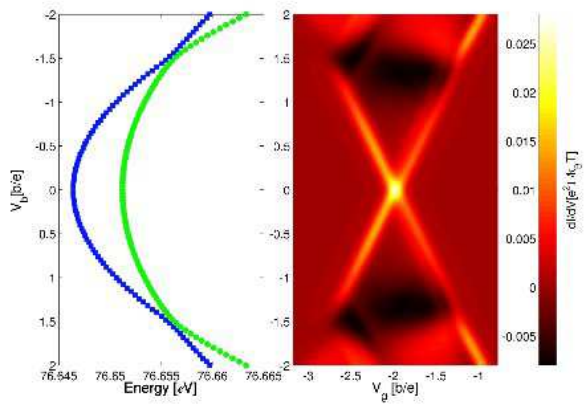}
\caption{(color online) Combination of the bias and contact
perturbations. Left panel - Energy levels of the 7 particle ground
and first excited state as functions of the bias voltage. Right
panel - Stability diagram around the $6\leftrightarrow7$ resonance. The
perturbation parameters are in this case $\xi_c = 2\Gamma$ and $a = 12$.}
  \label{fig9}
\end{figure}

Fig.~\ref{fig9} refers to the setup under both the \emph{bias and
contact perturbations}. The left panel shows the energy of the
lowest 7 particle states as a function of the bias. In the right
panel we present the stability diagram around the
$6\leftrightarrow7$ resonance. NDC and current blocking are
clearly visible only in the bias region where, due to the combination of
bias and contact perturbation, the two seven particle states return
quasi-degenerate. Also the fine structure in the NDC region is
understandable in terms of interference if in the condition of
quasi-degeneracy we take into account the renormalization of the
level splitting due to the energy non-conserving terms.

Interference effects predicted for the unperturbed benzene I-SET
are robust against various sources of symmetry breaking.
Quasi-degeneracy, $\delta E \ll  \hbar\Gamma \ll k_{\rm B} T$, is the necessary
condition required for the
detection of the interference in the stability diagram of the
benzene I-SET.

%///////////////////////////////////////////////////////////////////////////////

%///////////////////////////////////////////////////////////////////////////////
\section{Conclusions}\label{conclusions}
%///////////////////////////////////////////////////////////////////////////////

In this paper we analyze the transport characteristics of a
benzene I-SET. Two different setups are considered, the para
and the meta configuration, depending on the position of the leads
with respect to the molecule.

Within an effective $p_z$ orbital model, we diagonalize exactly
the Hamiltonian for the molecule. We further apply a
group theoretical method to classify the many-body molecular
eigenstates according to their symmetry and quasi-angular momentum. With the
help of this knowledge we detect the orbital degeneracy and, in the para
configuration, we select the states relevant for transport.

We introduce a generic interference condition \eqref{eq:interf_condition} for
I-SETs in terms of the tunnelling transitions \emph{amplitudes} of degenerate
states with respect to the source and drain lead. By applying it to the benzene
I-SET we predict the existence of interference effects in the meta
configuration

In order to study the dynamics of the molecular I-SET, we use a density matrix
approach which starts from the Liouville equation for the total
density operator and which enables the treatment of quasi-degenerate states.

The stability diagrams for the two configurations show striking
differences. In the linear regime a selective conductance
suppression is visible when changing from the para to the meta
configuration. Only transitions between ground states with well defined particle
number are affected by the change in the lead configuration. With the help of
the group theoretical classification of the states we recognize in
this effect a fingerprint of the \emph{destructive interference}
between orbitally degenerate states. We derive an analytical
formula for the conductance that reproduces exactly the numerical
result and supports their interpretation in terms of interference.
Other interference effects are also visible in the non-linear
regime where they give rise to NDC and current blocking at the border of
the 6 particle Coulomb diamond as well as to current suppression for
transitions between 7 and 6 particle states.

We provide a detailed discussion of the impact of the reduced
symmetry due to linking groups between the molecule and the leads
or to an electrostatic potential drop over the molecule. We classify different
transport regimes and set up the limits within which the discussed transport
features are robust against perturbations. We identify in the
\emph{quasi-degeneracy} of the molecular states the necessary
condition for interference effects.

\acknowledgments We acknowledge financial support by the DFG
within the research programs SPP 1243 and SFB 689.
%///////////////////////////////////////////////////////////////////////////////

%\newpage
\appendix

%///////////////////////////////////////////////////////////////////////////////
\section{GME and current in the non-secular approximation}\label{appendixGME}
%///////////////////////////////////////////////////////////////////////////////

The bias and the contact perturbations in our model for a benzene
I-SET lower the symmetry of the active part of the junction and
consequently lift the degeneracy that appeared so crucial for the
interference effects. The robustness of the latter relies on the
fact that the necessary condition is rather quasi-degeneracy,
expressed by the relation $\delta E \ll \hbar \G$.

Nevertheless, if the perfect degeneracy is violated, the secular
approximation applied to obtain Eq.~\eqref{eq:GME} does not
capture this softer condition. We report here the general
expression for the generalized master equation and the associated current
operator in the Born-Markov approximation and under the only further condition
(exact in absence of superconductors) that coherences between
states with different particle number are decoupled from the
populations and vanish exactly in the stationary limit:
\begin{widetext}
\begin{equation}
\begin{split}
 \dot{\s}^{\rm N}_{\rm E E'} = &-\frac{i}{\hbar}(E-E')\s^{\rm N}_{\rm E E'} +\\
 &-\sum_{\a\t\rm F}\frac{\G_\a}{2} \mathcal{P}_{\rm NE}
 \left\{
 d^\dagger_{\a\t}
    \left[
    -\frac{i}{\pi}p_{\a}(F-H^0_{\rm ben}) +
    f^-_\a(F-H^0_{\rm ben})
    \right]
 d^{\phd}_{\a\t} +
 \right.\\
 &\phantom{-\sum_{\a\t \rm F}\frac{\G_\a}{2} \mathcal{P}_{\rm NE}-}
 \left.
 d^{\phd}_{\a\t}
    \left[
    -\frac{i}{\pi}p_{\a}(H^0_{\rm ben}-F) +
    f^+_\a(H^0_{\rm ben}-F)
    \right]
 d^\dagger_{\a\t}
 \right\} \s^{\rm N}_{\rm FE'}\\
 &-\sum_{\a\t \rm F}\frac{\G_\a}{2} \s^{\rm N}_{\rm EF}
 \left\{
 d^\dagger_{\a\t}
    \left[
    +\frac{i}{\pi}p_{\a}(F-H^0_{\rm ben}) +
    f^-_\a(F-H^0_{\rm ben})
    \right]
 d^{\phd}_{\a\t} +
 \right.\\
 &\phantom{-\sum_{\a\t \rm F}\frac{\G_\a}{2} \mathcal{P}_{\rm NE}-}
 \left.
 d^{\phd}_{\a\t}
    \left[
    +\frac{i}{\pi}p_{\a}(H^0_{\rm ben}-F) +
    f^+_\a(H^0_{\rm ben}-F)
    \right]
 d^\dagger_{\a\t}
 \right\} \mathcal{P}_{\rm NE'}\\
 &+\sum_{\a\t \rm FF'}\frac{\G_\a}{2} \mathcal{P}_{\rm NE}
 \left\{
    d^\dagger_{\a\t}\s_{\rm FF'}^{\rm N-1}d^{\phd}_{\a\t}
    \left[
    +\frac{i}{\pi}p_\a(E'-F') + f^+_\a(E'-F') -
    \frac{i}{\pi}p_\a(E-F) + f^+_\a(E-F)
    \right] +
 \right.\\
 &\phantom{+\sum_{\a\t \rm FF'}\frac{\G_\a}{2} \mathcal{P}_{\rm NE}-}
 \left.
    d^{\phd}_{\a\t}\s_{\rm FF'}^{\rm N+1}d^\dagger_{\a\t}
    \left[
    +\frac{i}{\pi}p_\a(F'-E') + f^-_\a(F'-E') -
    \frac{i}{\pi}p_\a(F-E) + f^-_\a(F-E)
    \right]
 \right\}\mathcal{P}_{\rm NE'}
\end{split}
\label{gGME}
\end{equation}
\end{widetext}

\noindent where $\s^{\rm N}_{\rm EE'}$ is, differently to
Eq. \eqref{eq:GME}, in the Schr\"{o}dinger
picture. Eq.~\eqref{eq:GME} represents a special case of
Eq.~\eqref{gGME} in which all energy spacings between states with
the same particle number are either zero or much larger than the
level broadening $\hbar\G$. The problem of a master equation in presence of
quasi-degenerate states in order to study transport through molecules has been
recently addressed in the work of Schultz {\it et al.}\cite{Schultz09}.
The authors claim in their work that the singular coupling limit should be used
in order to derive an equation for the density matrix in presence of
quasi-degenerate states. Equation \eqref{gGME} is derived in the weak coupling
limit and bridges all the regimes as illustrated by Fig.~\ref{fig7}-\ref{fig9}.

The current operators associated to the master equation just presented read:

\begin{equation}
    \begin{split}
 I_\a &= \frac{\G_\a}{2} \sum_{\rm NEF\t}
 \mathcal{P}_{\rm NE}\\
 &\left\{\!
    \phantom{.}d^\dagger_{\a\t}
    \left[
        +\frac{i}{\pi}p_{\a}(E-H^0_{\rm ben}) +
        f^-_\a(E-H^0_{\rm ben})
    \right]
    d^{\phd}_{\a\t}
 \right.\\
 &+d^\dagger_{\a\t}
 \left[
    -\frac{i}{\pi}p_{\a}(F-H^0_{\rm ben}) +
    f^-_\a(F-H^0_{\rm ben})
 \right]
 d^{\phd}_{\a\t}\\
 &-d^{\phd}_{\a\t}
 \left[
    +\frac{i}{\pi}p_{\a}(H^0_{\rm ben}-E) +
    f^+_\a(H^0_{\rm ben}-E)
 \right]
 d^\dagger_{\a\t}\\
 &-\left.
    d^{\phd}_{\a\t}
    \left[
        -\frac{i}{\pi}p_{\a}(H^0_{\rm ben}-F) +
        f^+_\a(H^0_{\rm ben}-F)
    \right]
    d^\dagger_{\a\t}
 \right\}
 \mathcal{P}_{\rm NF}
    \end{split}
\end{equation}

where $\a = L,R$ indicates the left or right contact.
Nevertheless, within the limits of derivation of the master
equation, this formula can be simplified. Actually, if $E-F \leq
\hbar\G$, then $F$ can be safely substituted with $E$ in the
argument of the principal values and of the Fermi functions, with an
error of order $\frac{E-F}{k_{\rm B}T} < \frac{\hbar \G}{k_{\rm
B}T}$ which is negligible (the generalized master equation that we
are considering is valid for $\hbar \G \ll k_{\rm B}T$). The
approximation $E \sim F$ breaks down only if $E-F \sim k_{\rm
B}T$, but this implies $E-F \gg \hbar \G$ which is the regime of
validity of the secular approximation. Consequently, in this
regime, terms with $E \neq F$ do not contribute to the average
current because they vanish in the stationary density matrix.
Ultimately we can thus reduce the current operators to the simpler
form:
%
%\begin{widetext}
\begin{equation}
\begin{split}
 I_\a = \G_\a \sum_{\rm NE\t}
 \mathcal{P}_{\rm NE}
 \Big\{&
    +d^\dagger_{\a\t}
    \left[
        f^-_\a(E-H^0_{\rm ben})
    \right]
    d^{\phd}_{\a\t}\\
    &-
    d^{\phd}_{\a\t}
    \left[
        f^+_\a(H^0_{\rm ben}-E)
    \right]
    d^\dagger_{\a\t}
\Big\},
\end{split}
\end{equation}
which is almost equal to the current operator corresponding to the
secular approximation. The only difference is here the absence of
the second projector operator that allows contributions to the
current coming from coherences between different energy
eigenstates.
%///////////////////////////////////////////////////////////////////////////////

%///////////////////////////////////////////////////////////////////////////////
\section{Analytical formula for the linear conductance including the energy
non-conserving terms}\label{appendixPP}
%///////////////////////////////////////////////////////////////////////////////

In the derivation of the conductance formula \eqref{eq:G_NNp1} we
neglected the energy non-conserving terms in the
Eq.~\eqref{eq:GME}. Since in the GME they appear only in the
dynamics of the coherences between orbitally degenerate states,
Eq.~\eqref{eq:G_NNp1} is exact for the para configuration, where
the orbital degeneracy is cancelled. This is not the case in the
meta configuration where the orbital (quasi-)degeneracy is
essential for the description of interference. Thus we derived a
generic analytical formula for the conductance, taking into
account the energy non-conserving terms. It reads
\begin{widetext}
\begin{equation}
G_{\rm N,N+1}(\Delta E) = e^2\G \Lambda_{\rm N,N+1} \left[
    -\frac{S_{\rm N}S_{\rm N+1}f'(\Delta E)}
    {(S_{\rm N+1}-S_{\rm N})f(\Delta E)+S_{\rm N}}
\right] \left[
    1+\frac{{\rm aux}(S_{\rm N},S_{\rm N+1})
3\mathcal{P}^2 } {16 \Lambda_{\rm N,N+1}^2(S_{\rm N}S_{\rm N+1})^2
\left(f^{\pm}(\Delta E)\right)^{2} +\mathcal{P}^2}  \right].
\label{eq:Ggeneric}
\end{equation}
\end{widetext}
Here, it is $\G=\G_{\rm L}=\G_{\rm R}$. $\Lambda_{N,N+1}$ is the
overlap factor introduced in Section~\ref{LinCond_clean},
Eq.~\eqref{eq:lambda}. The auxiliary function ${\rm aux}(S_{\rm
N},S_{\rm N+1})$ in the correction term is zero if there are
\emph{no} orbitally degenerate ground states involved in the
transition. If one of the participating states is orbitally
degenerate it is ${\rm aux}(S_{\rm N},S_{\rm N+1})=1$. The sign in
$f^{\pm}(\Delta E)$ is defined as follows: $f^{+}(\Delta E)$ has
to be used if the $N$ particle ground state is orbitally
degenerate. If instead the $N+1$ particle ground state exhibits
orbital degeneracy, $f^{-}(\Delta E)$ has to be inserted. The
energy non-conserving terms are included in the factor
$\mathcal{P}=\mathcal{P}_{\rm L}\vert_{V_{\rm
bias}=0}=\mathcal{P}_{\rm R}\vert_{V_{\rm bias}=0}$. It is defined
only if a degenerate state is participating transport. In case
that {\it e.g.} the $N$ particle ground state is orbitally degenerate,
$\mathcal{P}_\alpha$ with $\alpha=L,R$ reads
\begin{widetext}
\begin{equation}
\begin{split}
\label{eq:PP} \mathcal{P}_{\alpha}=
 \sum_{{\rm E}^\prime,l}&\sum_{nm}
 \left[\frac{i}{\pi}p_{\alpha}(E_{\rm g,N}-E^\prime)\right]
 \bra{N-1,E^\prime\,l}d^{\phd}_{\alpha\tau}\ket{N_{\rm g},n}
 \bra{N_{\rm g},m}d^{\dagger}_{\alpha\tau}\ket{N-1,E^\prime\,l}\\
 \nonumber &-\sum_{{\rm E}^\prime,l}\sum_{nm}
 \left[\frac{i}{\pi}p_{\alpha}(E^\prime-E_{\rm g,N})\right]
 \bra{N+1,E^\prime\,l}d^{\dagger}_{\alpha\tau}\ket{N_{\rm g},n}
 \bra{N_{\rm g},m}d^{\phd}_{\alpha\tau}\ket{N+1,E^\prime\,l},
\end{split}
\end{equation}
\end{widetext}
where $p_\alpha(x)=-{\rm Re}\psi\left[\tfrac{1}{2} +
\tfrac{i\beta}{2\pi}(x -\mu_\alpha)\right]$ and $\psi$ is the
digamma function, as defined in Section~\ref{dynamics}.
%///////////////////////////////////////////////////////////////////////////////

%///////////////////////////////////////////////////////////////////////////////
\section{Transition probabilities for the $6\leftrightarrow7$
transition}\label{appendixC}
%///////////////////////////////////////////////////////////////////////////////

In the calculation of the overlap factor $\Lambda$ in
Section~\ref{LinCond_clean} we used the relation
\begin{equation}
|\bra{6_{\rm g}}d_{\rm L}\ket{7_{\rm g},\l=2}|^2 = |\bra{6_{\rm
g}}d_{\rm L}\ket{7_{\rm g},\l=-2}|^2.
\end{equation}
for the transition probabilities between the 6 particle ground
state and the 7 particle ground states $\ket{7_{\rm g},\l}$, where
$\l$ is the eigenvalue of the quasi-angular momentum. This
relation is
now to be proved.\\
Again, we take advantage of the symmetry properties of the
molecular states with respect to the $\s_{\rm v}$ operation and to
the rotation operator $R_{\phi}$ for rotations about a discrete
angle $\phi=\frac{n\pi}{3}$, as introduced in
Section~\ref{Symmetry}. The starting point is the generic relation
between these two operators:
\begin{equation}
\mathcal{R}_{\phi}\s_{\rm v}=\s_{\rm v}\mathcal{R}_{-\phi}.
\end{equation}
We can now apply both sides of this relation to the 7 particle
ground states $\ket{7_{\rm g},\l=\pm2}$:
\begin{equation}
\mathcal{R}_{\phi}\s_{\rm v}\ket{7_{\rm g},\l=\pm2}=\s_{\rm
v}\mathcal{R}_{-\phi}\ket{7_{\rm g},\l=\pm2}.
\label{eq:rot_refl_rel}
\end{equation}
The 7 particle ground states $\ket{7_{\rm g},\l=\pm2}$ are
eigenstates of each $\mathcal{R}_{\phi}$, and the corresponding
eigenvalues are phase factors:
\begin{equation}
 \mathcal{R}_{\phi} \ket{7_{\rm g},\l=\pm2}=
 e^{\mp2\cdot i\phi}\ket{7_{\rm g},\l=\pm2}. \label{eq:rot_eval_eq}
\end{equation}
Thus, Eq.~\eqref{eq:rot_refl_rel} becomes
\begin{equation}
 \mathcal{R}_{\phi}
 \Big(\s_{\rm v}\ket{7_{\rm g},\l=\pm2}\Big)=
 e^{\pm2\cdot i\phi}
 \Big(\s_{\rm v}\ket{7_{\rm g},\l=\pm2}\Big).
\end{equation}
Yet, according to Eq.~\eqref{eq:rot_eval_eq}, this equation can
only be valid if
\begin{equation}
\s_{\rm v}\ket{7_{\rm g},\l=\pm2}=\lambda\ket{7_{\rm g},\l=\mp2}.
\end{equation}
and, since $\sigma_{\rm v}^2 = 1$, $\lambda$ can only be a phase factor.
For the calculation of the transition probabilities we use further
the property $\s_{\rm v}^\dagger\s_{\rm v}^{\phd}=1$.
Since the left contact atom (atom 1) lays in the reflection plane
$\s_{\rm v}$, it is: $\s_{\rm v}^{\phd}d_{\rm L}\s_{\rm
v}^\dagger=d_{\rm L}$. Also, since the symmetry of the 6 particle
ground state is $A_{1g}$, it is: $\s_{\rm v}\ket{6_{\rm
g}}=\ket{6_{\rm g}}$. Under these considerations, we can write for
the transition probability to the state $\ket{7_{\rm g},\l=2}$:
\begin{eqnarray}
|\bra{6_{\rm g}}d_{\rm L}\ket{7_{\rm g},\l=2}|^2 &=& |\bra{6_{\rm
g}}\s_{\rm v}^\dagger\s_{\rm v}^{\phd}d_{\rm L}
\s_{\rm v}^\dagger\s_{\rm v}^{\phd}\ket{7_{\rm g},\l=2}|^2=\nonumber\\
&=&|\bra{6_{\rm g}}d_{\rm L}\s_{\rm v}\ket{7_{\rm g},\l=2}|^2=\nonumber\\
&=&|\bra{6_{\rm g}}d_{\rm L}\ket{7_{\rm g},\l=-2}|^2.\nonumber\\
\end{eqnarray}
%///////////////////////////////////////////////////////////////////////////////

%///////////////////////////////////////////////////////////////////////////////

%///////////////////////////////////////////////////////////////////////////////

\end{document}